\documentclass[3p,preprint]{elsarticle}
\usepackage{amssymb,amsmath}
\usepackage{graphicx,color}
\usepackage{bm}
\usepackage{xcolor}
\usepackage{gensymb}
\usepackage{subfigure}
\usepackage{verbatim}
\usepackage{listings}
\usepackage{graphicx}

\usepackage{ulem}

\begin{document}

\begin{frontmatter}
 \title{Analyzing Molecular Simulations Trajectories Using GPUs}
 \author[rvt]{Gourav Shrivastav\corref{cor1}}
 \ead{gourav.shri7@gmail.com}
 \author[rvt,els]{Manish Agarwal\corref{cor2}}
 \ead{zmanish@cc.iitd.ac.in, Tel. (+)91-11-2659-7170}
\cortext[cor1]{Corresponding author}
 \cortext[cor2]{Principal corresponding author}

 \address[rvt]{Department of Chemistry, Indian Institute of Technology Delhi, Hauz Khas, New Delhi 110016, Delhi, India}
 \address[els]{Current address: Computer Services Centre, Indian Institute of Technology Delhi, Hauz  Khas, New Delhi 110016, Delhi, India}
\date{\today}

\begin{abstract}
With the advent of high-performance computing techniques, the data for post-simulation analysis
has grown significantly. Depending upon the generated data size and the property of interest, the 
conventional serial codes may take few days for more compute intensive properties.  Here, graphic processing 
unit (GPU) based program kernels are discussed to exploit parallelism in the analysis codes specific to molecular 
simulations trajectories and data, hence reducing the time consumption. Commonly targeted properties 
of systems which are based on static as well as dynamic correlations are considered. Static properties considered 
here include bond order parameter, tetrahedral order parameter, and triplet
correlation functions, in order of increasing work load as described in
conventional serial logic. In dynamic properties, 
the auto-correlation function, involved in Green-Kubo method, and the mean square displacement, as per Einstein 
method, calculations are discussed. Particularly, for these embarrassingly parallel and computationally intensive 
problems, the compute unified device architecture (CUDA) application programming
interface (API) affords considerable speed up 
with minimum implementation effort. Depending upon the system size, the computation time of static properties is 
shown to be reduced by 10-50$\times$ relative to conventional serial CPU codes. A similar speed up of 
10-80$\times$, depending on the computation parameters, is also achieved for dynamic 
properties. Since these GPU codes render 1-2 order of magnitude reduction in the duration of analyses, these 
codes, in conjunction with GPU accelerated molecular simulations may lead to an overall improved and efficient 
performance. The methodology discussed here is very generic and may be extended to more specific and complex 
analysis codes.
\end{abstract}

\end{frontmatter}

\section{Introduction}
Recent emergence of high-performance computing (HPC) techniques and relatively
easy access to large clusters has resulted in development and use of
computational methods as a recurrent tool in many interdisciplinary scientific
research areas such as quantum mechanics\cite{ky08,ky08a,um08}, molecular
mechanics\cite{alt08,lsvm08,stc10,smag17}, molecular modelling\cite{adof16},
medical imaging\cite{shus10}, and cosmology\cite{pmvss12,bbayk13}. At the core of these HPC
techniques, the many core/thread architecture of Nvidia GPUs cards render a pivotal 
role to harness parallelism while keeping the economic and environmental footprint low. More specific to molecular
dynamics (MD) simulation studies, various modern simulation packages
LAMMPS\cite{sp95,bwpt11,bkpt12}, GROMACS\cite{slhgmb05}, DL\_POLY\cite{tstd06},
NAMD\cite{pbwgtv05}, CHARMM\cite{bbmnpr09}, Gaussian\cite{g16} etc. are
utilizing these technologies for compute-intensive calculations, such as
generation of neighbour lists and force calculations. MD simulation packages, like HOOMD\cite{alt08}, RUMD\cite{bihvb15}, and ACEMD\cite{hgf09}, are also available which perform entire calculations on 
GPUs. In addition to HPC resources, low to mid range consumer GPUs have enough compute power and 
memory to accelerate simulations 2-3 times faster as opposed to that on only CPUs albeit with a limitation of memory. Consequently, massive size
simulations, e.g. complex biomolecular compounds or multi nanoparticle systems
under solvent conditions, have now become tractable on workstation level
machines. With these advancements, the length and time scale scope of molecular
simulations has grown nearly exponentially in the past decade. Length scales upto 10 to 100nm are now accessible and multi billion atom simulations have been performed\cite{skrh18}. With coarse graining approaches such as MARTINI for simulation of full viruses with explicit solvent,  access to longer time scales and multi-$\mu$s simulations have been reported\cite{kwbm18,be18}. Hence, the data generated and required calculations for 
analyses also have grown in tandem. Although the growth in capabilities of simulation packages is remarkable 
and evident from literature, GPU-based analysis codes of the resultant data are not easily
available, with few exceptions\cite{alt08,lsk11}. Some of the above mentioned packages also offer analysis codes which are well optimized to run on modern CPU architectures.    

In molecular simulations, some of the analyses can be made \textit{on-the-fly}.
However, it is often preferred that the raw data is obtained, so as to
extract more information post-simulation than provided by the package during the lifetime of the simulation. 
Standard texts in the field describe serial algorithms used for analysing molecular dynamics data, emphasising clarity and correctness, leaving it to the reader to explore methods to speed up the calculations\cite{fs02,at87}. 

Recently, time taken for running of some complex {\it serial} analysis codes has reached similar or more 
than the time taken 
by the simulation itself, instigating a necessity to optimize the algorithms for modern architectures, and/or port these analysis codes on parallel architectures to gain
overall performance.  There are many shared memory and distributed memory parallelization application 
programming standards and interfaces available. In terms of increasing ease of coding some of these are 
Message Passing Interface (MPI), Common Unified Device Architecture (CUDA), Open Multi-Processing 
(OpenMP) and Open Accelerators (OpenACC). Distributed memory APIs such as MPI involve considerable changes in the code since copies of the full code run simultaneously, and usually need to inter-communicate. Compiler directive driven standards such as OpenMP and OpenACC rely on the compiler's capabilities for parallelization. Furthermore, use of OpenACC is currently restricted to Portland Group compilers, limiting its accessibility. CUDA, on the other hand, affords a good balance between programmer control and process speed up. This also involves considerable change in the original code, however, small portions of the code, for example data parallel functions or loops can be ``parallelized''. In this work, CUDA based GPU-accelerated versions of common analyses for MD simulation trajectories are presented, and are compared to corresponding naive CPU codes. The authors note that implementation of well optimized codes in all these ansatz, whether MPI, OpenMP/OpenACC or CUDA would take similar amount of time and effort, and yield similar efficiencies. 

MD simulations give access to both static and dynamic properties of the system.
The radial distribution function, $g(r)$, is one of the most common algorithms and has
been utilized in many areas of scientific research\cite{sk93,kglh02,vkknv04,igcf10}, and has been well parallelized
and optimized\cite{lsk11}. The calculation of $g(r)$ involves the compute-intensive, but
extremely parallel search for pair distances and depends
quadratically on the system size. Other order parameters and static properties
also often depend on neighbour searches. In the present study, two static properties, tetrahedral order parameter 
($q_{tet}$)\cite{ch98} and crystalline order parameter ($P_2$)\cite{ylr13}, are discussed to exploit parallelism for analysis of trajectory data obtained from molecular dynamics simulations of systems such as water and polyethylene respectively. Both these order parameters utilize similar core structure of
calculations as that of radial distribution functions (RDFs), and therefore, have a highly parallelizable
algorithm. Another investigated static property is triplet correlation functions or three-particle correlation 
functions (TPCFs)\cite{be89}, an important quantity in systems where two-body correlations are not sufficient 
to probe correlations \cite{sdnmc14}. TPCFs extend the calculation of pair distances to an additional level and, therefore, 
can reveal information about the angle resolved local ordering. It first calculates all the possible pair distances, 
like in the case of RDFs, subsequently, distances of all the particles are calculated for each pair.  Hence, analysis time varies as $n^3$ with number of particles, $n$, and makes it a compute-expensive analysis scheme. Among
dynamic properties, the computation of viscosity and diffusivity \cite{fs02} is discussed using the Green-Kubo 
and the Einstein formalism, respectively. The Green-Kubo method evaluates the viscosity as the time integral 
of off-diagonal terms of pressure tensor.  The calculation of diffusivity via Einstein method is based on the 
average mean square distance travelled by particles in a given time, and like time-correlation functions, it
utilizes all the possible time origins.  Since these algorithms have inherent data parallelism,  
CUDA on GPUs is an appropriate choice. It
must be noted that although rigorous optimization of such code can yield
extremely high speed ups, it comes at the cost of learning fine details of
architecture and/or APIs, as well as time taken to achieve the same. The motivation for the current work is to provide templates for standard analysis approaches with considerable speed up from minimal knowledge, learning time and implementing the parallel 
code, and thereby, to enhance overall productivity of molecular simulations
by reducing the compute load of analysis. An ancillary effect might be promotion
of ideas of parallelism and use of existing resources to beginners, which is
currently limited to relative few career programmers or enthusiasts. The work has been aimed at audience who have traditionally written codes for analysis with an aim of correctness, and little or no experience in code optimization. Hence, the analysis time recorded from {\it unoptimized} GPU codes is compared with those obtained from typical serial unoptimized CPU codes.

In summary, the present work covers two basic approaches for analysis of data obtained from typical MD simulations, namely, spatial correlations based on distance based searches within a single configuration over a trajectory, and temporal correlations based on time analysis. Tetrahedral order and polymer crystalline order parameters described above cover basic pair searches in non bonded and bonded systems respectively. Triplet distribution $g^{(3)}$ describes searching of triads under given constraints. It is also an example where the {\it serial} analysis code takes longer to run than the well parallelized simulation. ACF and MSD calculations cover the Green-Kubo and Einstein approaches for calculation of micro and bulk dynamic properties of the system. The algorithms and codes are sufficiently modular and can be adapted to any related property. 

The paper is organized as follows. Section 2 contains computational details which includes simulation details, hardware details, and introduction to a CUDA kernel.  Section 3 provides discussion on the parallel pseudo codes of utilized algorithms along with the results obtained. Conclusions are given in Section 4.

\section{Computational and Simulation Details} 
In order to describe the parallelization of the common analysis codes, we use the following simulations of representative systems.

\subsection{Simulation Details} 
Three varieties of systems were considered for this study, water, $n$-octane, and polyethylene.
To model water, the three site single point charge (extended) (SPC/E) model was used \cite{bgs87}. For 
$n$-octane and polyethylene, the united atom version of TraPPE\cite{ms98,cte10,sack17} and all atom version 
of OPLS force-fields\cite{rmj96} were used. For both water and $n$-octane, simulations were performed in stable 
liquid regime using LAMMPS\cite{sp95,bwpt11,bkpt12} MD simulation package to obtain the trajectories.
The equations of motion were integrated using the velocity-Verlet scheme with a time step of 1 fs. Orthogonal 
periodic boundary conditions were employed to ensure the bulk limit and a global cut-off of 14 $\mathring A$ was used for 
short range interactions. Temperature and pressure were maintained using Nos{\'e}-Hoover thermostat and 
barostat, respectively. Particle-particle-particle mesh was used for the k-space calculations. Rigid constraints 
associated with water molecules were maintained by the use of SHAKE algorithm. For both systems, 
initial random configuration consisted 6720 molecules was obtained by PACKMOL\cite{mabm09}. These 
configurations were then subjected to minimization followed by 2 ns long equilibration in isothermal and 
isobaric ensemble. Temperature and pressure were maintained at 300 K and 1 atm, respectively. The resulting 
configuration was then replicated to obtain the configurations for the system containing 13440 and 26880 molecules.
Subsequently, systems were allowed to simulate in canonical ensemble in  Trajectories were saved for every 100
steps. For polyethylene, data from our previous study for the system containing 40 chains of 50mers, i.e., 2000 
carbon atoms with 2000 frames was used \cite{sa16}.

\subsection{Hardware Details and Programming Environment}
For detailed testing and deploying analysis codes, we have used two kinds of machines with K40m 
(2880 cores, 12 GB memory, 0.745 GHz clock rate) and GTX 780 (2304
cores, 3 GB memory, 0.993 GHz clock rate) GPUs. The K40 machine contains
Intel(R) Xeon(R) CPU E5-2680 v3 with the clock rate of 2.50 GHz and 64 GB
memory while the GTX machine contains Intel(R) Xeon(R) CPU E5-1650 v2 with
clock rate of 3.50 GHz and 32 GB memory. All MD simulations were performed on the
K40 machine. Intel compilers (v15) with default flags were used to compile all
serial codes. CUDA v6.5, along with GNU C and FORTRAN compilers (v4.4), was used for compiling all accelerated codes. Since these GPUs are considerably old, the analysis codes were also tested on a limited scale on comparatively recent P100 and V100 GPUs with CUDA 8.5 and 9.0 respectively. Furthermore, the K40 machine is part of a cluster with a high performance lustre parallel file system. Considering that the total data being used for benchmark purposes is small, the bottlenecks were the calculations themselves, and not the reading of data. Where ever relevant, time taken to read the data has been subtracted from the benchmark time. It is possible to further amortize the cost of data reading, as well as transfer to GPU memory, i.e. ``hide'' the data latency, but this discussion is beyond the scope of this work.

\subsection{CUDA Kernels}
CUDA is a programming paradigm introduced by Nvidia to leverage parallelism on multi-threaded GPU devices.
The toolkit also includes highly optimized libraries, of varied algorithms for
scientific computations, 
such as fast Fourier transform library, sparse matrix library, etc. A conventional CUDA program accommodates 
both serial and parallel structure.  In this context, CUDA may utilize both C/C++ and FORTRAN languages. 
In general, serial part executes on CPU and the parallel part, also known as CUDA kernel, is ``offloaded'' to the
GPU. For the processing of data, CUDA implements single instruction multiple thread (SIMT) model which implies 
that every thread in a block executes a single instruction on different data. CUDA hides memory latency in order to
improve parallel performance.

Pseudo code~\ref{al:CUDA} is a representative CUDA program for computing a specified parameter. 
Here, we have discussed the mandatory components of a parallel CUDA program.  At first, a kernel 
`CUDA\_parameter' is defined which utilizes the input to update `d\_parameter' (line 4-14) followed by the main program.  The main program defines the required variables (line 19-25), allocates
memory on host and device appropriately (line 29-36),  and read input files
(line 45-46). Subsequently, the data is transferred to GPU device (line 49) and
a kernel for parallel execution is launched using `nblock' blocks with
`nthread' threads each (line 54). Once the calculation in kernel is over, data
is copied back from device to host (line 64) for further processing and the allocated memory is freed
(line 69-74).

\begin{lstlisting}[frame=lines, language=C++, caption={A representative CUDA
kernel program}, label={al:CUDA}]
//include required files

using namespace std;
//Define kernel
__global__ void CUDA_parameter(double *d_x, double *d_y, double *d_z,
double *d_parameter,double xbox, double ybox, double zbox, int Natom)
{
 int i;
 i=blockId.x*blockDim.x+thread_id.x; // atom index is the thread-id 
 if ( i < Natom ) {
   //Compute parameter
   //update d_parameter
 }
}

//Define main program
void main(){
    //define device variables 
    double *d_x,*d_y,*d_z;
    double *d_dx,*d_dy,*d_dz,*d_dr;
    double *d_param;
    //define host variables 
    double *h_x,*h_y,*h_z;
    double *h_param;
    int nthreads=128,nblock;

    //Read Natom = number of atoms

     unsigned long long int sized=Natom*sizeof(double);

     //allocate device memory e.g.:
     cudaMalloc((void**)&d_x,sized);
     //allocate other variables

     //allocate host memory e.g.
     h_x= (double *) malloc(sized);
     //allocate other variables

     //Check for valid allocation e.g. cudaGetLastError()
     
     //compute number of blocks
     nblock=(int(numatm/nthreads)+1);
    

     //Start loop(1) to read data from file
     //e.g. read data for one frame

     //transfer data from device to host e.g:
     cudaMemcpy(d_x,h_x,sized,cudaMemcpyHostToDevice);
     //transfer other data


     //invoke kernel
     CUDA_parameter<<<nblock,nthreads>>>
     (d_x,d_y,d_z,
     d_parameter,xbox,ybox,zbox,Natom);
     
     //ensure kernel has finished executing
     cudaDeviceSynchronize();

     //loop(1) end

     //transfer data from host to device e.g:
     cudaMemcpy(h_param,d_param,sized,cudaMemcpyDeviceToHost);

     //Write data to file or process further

     //free device memory e.g.:
     cudaFree(d_x);
     //free other memory
     //free host memory e.g.:
     free(h_x);
     //free other memory
  }
}
\end{lstlisting}

Since C and FORTRAN have different ways of storing two dimensional arrays, we use linear arrays extensively all throughout the codes, i.e., two dimensional arrays are mapped onto a one dimensional array. 

\section{Results and Discussions}
Prior to comparing the performance achieved from GPUs, it is confirmed that (i)
the computed quantities are in accordance with the reported values in
literature and (ii) the outputs obtained from GPU programs are equivalent to
those obtained from conventional CPU programs. All codes are maintained via public repository ``GPU-MD-Analysis/MDGPUAnalysis'' on github.com  (https://github.com/GPU-MD-Analysis/MDGPUAnalysis.git). 

\subsection{Two Particle Parameters}
The RDF or pair correlation function is one of the most common static analysis schemes for molecular 
simulations trajectories. For a given trajectory frame, RDFs give probability of finding a neighbour at a 
particular distance by determining all the possible pair separations. Since the data from a single frame is not
sufficient to satisfy statistical accuracy for common simulation sizes, an additional loop over the number of frames is also required. 
Note that calculations in a given trajectory frame are independent of the rest of trajectory frames. 
Furthermore, the calculation of all the pair separation in a given frame can also be performed 
independently. Since RDF has been discussed before in great detail\cite{lsk11}, we do not describe it here. 
However, both the considered order metrics, $q_{tet}$ and $P_2$, have a common algorithmic
feature of finding pair distances, a nested loop over the number of particles,
as in the case of RDF. The only difference is in final processing, where
resulting pair distances are used appropriately.
It clearly conveys that similar parallel and logical structures can be utilized
to perform accelerated computations for these metrics. Note that
the time taken for such calculations will also depend on the state point, i.e.
density of the system. The time taken can be as high as
30\% in some cases, for {\it both} CPU and GPU algorithms. While the additional
cost of calculation per particle is {\it accumulated} in the CPU code, the
same is amortized over the parallel CUDA cores, hence resulting in improved
speed ups. 

\subsubsection{Tetrahedral Order Parameter}
Tetrahedral fluids belong to a class of complex fluids which are known to form
networks\cite{dc13}. The $q_{tet}$ parameter has been extensively used to
understand the local orderings in water molecules and other tetrahedral fluids, 
viz. beryllium fluoride and silica\cite{ac07,ac09,assac10}. In such systems, $q_{tet}$ is used to
quantify the extent to which particles are organized in a tetrahedral manner.
Mathematically, for a single particle, it is given as \cite{ch98,ed01}

\begin{equation}
\label{eq:qtet}
q_{tet}=1-\frac{3}{8}\sum_{j=1}^{3}\sum_{k=j+1}^{4} (\cos\psi_{jk} +1/3)^2,	
\end{equation}

where $\psi_{jk}$ is the angle between the bond vectors $r_{ij}$ and $r_{ik}$,
where $j$ and $k$ label the four nearest neighbour atoms of the same type in
the vicinity of atom $i$. The pseudocode (FORTRAN) of a conventional program for calculation of $q_{tet}$ is given in Algorithm~\ref{al:qtet}.

\begin{lstlisting}[frame=lines, language=Fortran, mathescape,
caption={Serial logic for tetrahedral order parameter},label={al:qtet}]
Allocate memory for array of positions on CPU
Allocate memory for arrays of neighbour vectors and qtet on CPU
Do frame = 1 to Nframe    ! Loop 1 over Nframe frames
   Do i = 1 to  Natom          ! Loop 2 over Natom particles
      Do j = 1 to Natom        ! Loop 3 over Natom particles
        Calculate the minimum image distance vector between particle i and j
	store all distances
      End Do                         ! End of Loop 3 
   Sort the distances to get the four nearest neighbours
      Do j = 1 to 3                 ! Loop 4 
         Do k = (j+1) to 4       ! Loop 5  
          Calculate the $\cos \psi_{jk}$
          Solve Eq.$~\ref{eq:qtet}$
         End Do                     ! End of Loop 5  
      End Do                        ! End of Loop 4  
      store $q_{tet}$
   End Do                           ! End of Loop 2  
   Update histogram with $q_{tet}$
End Do                              ! End of Loop 1 
Free memory 
\end{lstlisting}

As shown in Algorithm~\ref{al:qtet}, in order to calculate $q_{tet}$ for a given atom $i$, we need to calculate 
all the neighbour distances, followed by a sorting to get the four nearest neighbours. Subsequently, using each 
nearest neighbour pair ($j$ and $k$), $\angle jik$ (using Eq.~\ref{eq:qtet}) is calculated to get the value 
of $q_{tet}$ for each particle in each frame (Loop 2, Algorithm~\ref{al:qtet}) of the trajectory in a 
sequential manner. Overall, it depicts the complexity of ${\cal O}(Nframe(Natom)^2)$. Since $q_{tet}$ of 
every atom in each frame is a local quantity, parallel threads on GPU device can be launched for every pair 
of frame and atom. While small systems or small trajectory data can be benefited with this scheme, the same 
is not true for the bigger data size since the whole trajectory might not completely fit in GPU device memory. In order to counter the memory 
constraints for larger systems, alternatively, data only for a single frame can also be sent to the GPU. Loop 2
(Algorithm~\ref{al:qtet}) can be launched in parallel where each thread will
perform Loop 3 (Algorithm~\ref{al:qtet}) to compute $q_{tet}$ for one particle.
The execution on these threads is scheduled internally by the GPU architecture.
The pseudocode of parallel GPU program for the $q_{tet}$ is given in
Algorithm~\ref{al:qtetpar} 

\begin{lstlisting}[frame=lines, language=Fortran, caption={Parallel logic for tetrahedral order parameter},label={al:qtetpar}]
Allocate memory for arrays of positions and qtet on CPU and GPU 
Allocate memory for array of neighbour vectors on GPU  
Do frame = 1 to Nframe ! Loop 1 over Nframe frames 
     Transfer array of positions to GPU
     ! Kernel call. cf. Algorithm $\ref{al:CUDA}$ line 54-56
     Call CUDA_qtet over all atoms  
     Transfer qtet histogram to CPU
End Do                 ! End of Loop 1  
Free memory

function CUDA_qtet
If (thread_id < Natom) 
  Compute qtet on device for one atom by performing Loop 3, 4, 5 of Algorithm$~\ref{al:qtet}$
  Update qtet histogram using atomic operation e.g. atomicAdd
End function
\end{lstlisting}

It may also be noted that the calling program and the CUDA kernel ({\it
CUDA\_qtet}) are entirely independent and hence can be written in different
languages. We use FORTRAN for the former to read the trajectory and writing
outputs, while CUDA-C is used for the latter to compute $q_{tet}$. The speedups
gained from this approach of parallelizing over number of particles for
different system sizes and available GPU machines are shown in
Table~\ref{tab:qtet}. Typical speedups are $\approx20-50\times$ compared to the serial code. This illustrates that
considerable performance improvement can be achieved by minimal changes in the
program. Timing for the smallest system size for P100 and V100 cards are respectively 9.44 s and 6.4 s respectively for 100 frames, indicating that the algorithms scale well. Furthermore, time taken for 1000 frames is 75.35 s and 43.78 s, confirming that timing scales sub linearly with increasing data.  
\begin{table}[htbp]
\centering
\begin{tabular}{@{\extracolsep{4pt}}rrrrrr@{}}
\hline 
\hline 
\multicolumn{1}{c}{System Size} &\multicolumn{1}{c}{CPU}&\multicolumn{2}{c}{GTX} &\multicolumn{2}{c}{K40}\\ 
\cline{3-4}
\cline{5-6}
 & Time (s) & Time (s) & Speed up & Time (s) &Speed up\\
\hline 
 6720  &  184  &   7 &  26  & 8   & 23  \\ 
 13440 &  994  &  19 &  52  & 23  & 43  \\ 
 26880 &  3560 &  63 &  56  & 60  & 59  \\ 
\hline
\end{tabular}
\caption{Performance comparisons for computation of $q_{tet}$ using conventional 
serial CPU and parallel CUDA codes, on GTX and K40 machines. Computation 
is performed for 100 frames and relative speed up is also given. Note that time of 
file reading is not included in CPU time ($<$ 1 s).}
\label{tab:qtet}
\end{table}

\subsubsection{Bond Orientational Parameter}
Another order metric which also involves discovery of nearest neighbour pairs is $P_2$.
This local order parameter quantifies the extent to which chain segments are
aligned in the vicinity of a given particle and often used
for long chain compounds/polymers\cite{ylr13,sa16}. Mathematically, local $P_{2,i}$
order parameter of the $i$${^th}$ carbon atom in a chain is defined
as\cite{ylr13}
\begin{equation}
\label{eq:p2}
P_{2,i}=\frac{1}{n}\sum_{j=1}^{n}\frac{3\cos^2(\theta_{ij})-1}{2},
\end{equation}
where $\theta_{ij}$ is the angle between the vector from the $(i - 1)$${^th}$ to
the $(i + 1)$${^th}$ carbon atom and the vector from the $(j - 1)$${^th}$ to the $(j +
1)$${^th}$ carbon atom that lie within a certain cut-off distance. The value of
$P_2$ varies in the range of 1 to -0.5 for a perfect crystalline arrangements to
completely random configurations, respectively. The logical structure 
for the calculation of $P_2$ is described in Algorithm~\ref{al:p2}.

\begin{lstlisting}[frame=lines, language=Fortran, mathescape, caption={Bond order parameter}, label={al:p2}]
Allocate memory for arrays of positions and $P_2$ on CPU
Do frame = 1 to Nframe          ! Loop 1 over Nframe frames
   Do i = 1 to Natom            ! Loop 2 over Natom atoms
      If (carbon atom i is bonded to two carbon atoms)  ! If condition 1
      Do j = 1 to Natom         ! Loop 3 over Natom atoms 
         If (i.ne.j .and. j is bonded to two carbon atoms) ! If condition 2
           Calculate the minimum image distance between i and j 
           If (distance is within the cut-off)  ! If condition 3
             Find the vector $(i - 1)$th to the $(i + 1)$th atom
             Find the vector $(j - 1)$th to the $(j + 1)$th atom
             Solve dot product of vectors to get angle
             Solve the Eq.$~\ref{eq:p2}$
             Assign $P_2$ value to atom i
           End if                     ! End of If condition 3
         End if                       ! End of If condition 2
      End Do                          ! End of Loop 3 
      End if                          ! End of If condition 1 
   End Do                             ! End of Loop 2 
End Do                                ! End of Loop 1
Free memory
\end{lstlisting}

Just like in $q_{tet}$, it can be parallelized over the number of atoms in each
frame. However, in order to efficiently use the GPU architecture, additional
parallelism can be exploited over frames. To accomplish this goal, threads
are launched for each pair of frame-atom in a kernel call. This ensures that
time between the launch of subsequent kernels in the previous case of $q_{tet}$
is eliminated, and all data can be processed in a single kernel call, as
described in Algorithm~\ref{al:p2par}.  The caveat, of course, is that the
full data must fit on the GPU.  Since these are embarrassingly parallel
algorithms, data can always be {\it designed} to fit on a single device e.g.
a single trajectory can be split before processing and the results can be
merged afterward. 

\begin{lstlisting}[frame=lines, language=Fortran, mathescape, caption={ Parallel bond order parameter}, label={al:p2par}]

Allocate memory for arrays of positions and $P_2$ on GPU and CPU
Transfer all data to GPU
Call CUDA_p2 ! Calling parallel calculation for all the atoms of all frames
Transfer results to CPU
Free memory

function CUDA_p2
If (thread_id < Nframe*Natom)                ! If condition 1
	Frame index = int(thread_id/Natom)           ! Fetching frame index $\label{line:fetch}$
	Atom index = (thread_id %Natom)              ! Fetching atom index for atom i
	If (carbon atom i is bonded to two carbon atoms)  ! If condition 2
		Compute $P_2$ value for each carbon using Loop 3 of Algorithm$~\ref{al:p2}$
	End If
End If      
End function
\end{lstlisting}
Using these strategies for a system of 2000 carbons\cite{sa16} with 2000
configurations in the trajectory, serial CPU computation takes ~162 seconds.
On the other hand, the parallel computation on K40 and GTX machines takes 39
and 14 seconds, respectively. It is not surprising that the speed up achieved is
less ($\approx4\times$) since the calculation is itself very small. Better speed
up is expected with the large number of atoms and frames. 

\subsection{Three-Particle Parameter or Triplet Correlations}
Three-particle correlation functions (TPCFs),
$g^{(3)}(\mathbf{r1},\mathbf{r2},\mathbf{r3})$, can be described as the
probability of finding three particles simultaneously at positions
$\mathbf{r1}$, $\mathbf{r2}$, and $\mathbf{r3}$ in the system, irrespective
of the configuration of the remaining ($N - 3$) particles
\cite{sgrdk17,sack17}. In general, it can be given as\cite{dam04} 
\begin{equation}
  g^{(3)}({\bf r_1,r_2,r_3})=\frac{N(N-1)(N-2)}{\rho^3Z_N} \int \exp(-\beta
  U({\bf r^N}))d{\bf r^{(N-3)}} ,
  \label{eq:g3}
\end{equation}
where $\rho$, $Z_N$ and $U({\bf r^N})$ are total number density, the configuration integral, and interaction 
potential with $\beta=1/k_BT$, respectively. Unlike RDFs, which gives radial probability averaging over all the 
angular possibilities, TPCFs can be used to obtain angle-resolved information. The computation of $g^{(3)}$ 
involves 3D histogram binning of distances with imposed conditions. The detailed information about the implication
of such conditions are discussed in Refs~\cite{mmhr83,sdnmc14}. However, one of such conditions on triplet is that
the two of its distance should be less than $1/4^{th}$ of box length and the third distance should be less than 1/2 of box length. 
In comparison to two-particle correlations, TPCFs involve an additional loop over the number of particles which 
significantly reduces the performance and increases the overall analysis time. The serial logic for TPCFs is shown in
Algorithm~\ref{al:g3}.

\begin{lstlisting}[frame=lines, language=Fortran, mathescape, caption={Serial logic for triplet correlation function}, label={al:g3}]
Allocate memory for arrays of positions, updated positions on CPU
Allocate memory for g2 and g3 histograms on CPU 
Do frame = 1 to Nframe                 ! Loop 1 over Nframe frames
   Do i = 1 to Natom                   ! Loop 2 over Natom atoms
      Do j = (i+1) to Natom            ! Loop 3 
         Calculate minimum image distance between atom i and j
         Compute and store new coordinates of neighbour particles
         Assign two particle histogram
      End Do                            ! End of Loop 3
      Do j = 1 to Natom-1               ! Loop 4 over Neighbour particles
         If (distance of i and j is within Boxlength/4 )       ! If condition 1
         Do k = (j+1) to Natom-1        ! Loop 5 
            If (distance of i and k is within Boxlength/4 )    ! If condition 2
              Calculate the distance using updated positions
              If (distance of j and k is within Boxlength/2 )  ! If condition 3
                Update g3 histogram
              End If                    ! End of if condition 3
            End If                      ! End of if condition 2
          End If                        ! End of if condition 1
         End Do                         ! End of loop 5
       End Do                           ! End of loop 4
    End Do                              ! End of loop 2
End Do                                  ! End of loop 1
Free memory
\end{lstlisting}
In the case of the corresponding CUDA program, a slightly different logic is used.  First, pair
histograms and pair vectors are computed in parallel for all particles in a
configuration. Subsequently, kernels are launched for each particle pair
asynchronously. Each thread computes and updates the global $g^{(3)}$ 3D
histogram, as shown in Algorithm~\ref{al:g3par}. {\it n.b. } g2 histogram is computed with cut-off of 1/2 of box length while
neighbour vectors are stored within a cut-off of $1/4^{th}$ of box length.
Note that while computing g3 histogram, the neighbour vectors computed in kernel CUDA\_g2
(Algorithm:~\ref{al:g3par}) are utilized.

\begin{lstlisting}[frame=lines, language=Fortran, mathescape, caption={Parallel logic for triplet correlation function}, label={al:g3par}]
Allocate memory for arrays of positions on GPU and CPU
Allocate memory for arrays of neighbour vectors on GPU only
Allocate memory for g2 and g3 histograms on CPU and GPU 
Do frame = 1 to Nframe     ! Loop 1 over Nframe frames
  Transfer data for this frame to GPU
  Call CUDA_g2             ! Calling parallel calculations over Natom
  Do i = 1 to Natom        ! Loop 2 
    Call CUDA_g3           ! Calling parallel calculation over neighbour vectors    
  End Do                   ! End of loop 2
End Do                     ! End of loop 1
Transfer g2 and g3 histogram to CPU
Free memory 

function CUDA_g2
  If (thread_id < Natom)   ! If condition 1
  Do j = 1 to Natom        ! Loop 3 over Natom atoms
  Compute and store g2 histogram for atom i
  Compute and store minimum image neighbour vectors for atom i.
  End Do                   ! End of Loop 3
  End If                   ! End of If condition 1
End function

function CUDA_g3
  Identify current neighbour j of atom i as thread_id
  Do k = 1 to number of neighbours of i  ! Loop 4 over neighbours
     Compute j-k distance                 
     If distance j-k < boxlength/2       ! If condition 2           
        Update g3 histogram
     End If                              ! End of If condition 2
  End Do                                 ! End of Loop 4 
End function
\end{lstlisting}

Table~\ref{tab:g3} shows the time consumed for the computation of TPCFs. We
note here that serial computation of $g^{(3)}$ is extremely slow. The computations usually
requires large system sizes as well as a long trajectory for convergence and
good ensemble averages. For example, the serial code (unoptimized) takes
$\approx$3~hours to compute $g^{(3)}$ for 10 frames in the case of
largest system size considered here (Table~\ref{tab:g3}). In contrast, the same
computation on a consumer GTX780 takes $\approx$13$\times$ less time, i.e.
$\approx$14 {\it minutes}.  
\begin{table}[htbp]
\centering
\begin{tabular}{@{\extracolsep{4pt}}rrrrrr@{}}
\hline 
\hline 
\multicolumn{1}{c}{System Size} & \multicolumn{1}{c}{CPU} & \multicolumn{2}{c}{GTX} & \multicolumn{2}{c}{K40} \\ 
\cline{3-4}
\cline{5-6}
 & Time (s) & Time (s) & Speed up & Time (s) &Speed up\\
\hline 
  6720  &  596  & 13  & 45 & 16   & 37 \\ 
 13440  & 2089  & 104 & 20 & 132  & 16 \\ 
 26880  &11515  & 860 & 13 & 1124 & 10 \\ 
\hline
\end{tabular}
\caption{Performance comparisons for computation of $g^{(3)}$ using conventional 
serial CPU and parallel CUDA codes, on GTX and K40 machines. Computation 
is performed for 10 frames and relative speed up is given. Time taken for file
reading is included in the data. Time taken for writing output is 2-3s on the GTX machine
and 12-14s on the K40 machine, not included in the data above.}
\label{tab:g3}
\end{table}
\subsection{Correlation functions}
\subsubsection{Viscosity}
In equilibrium molecular dynamics, viscosity can be computed by utilizing
Green-Kubo method. This method utilizes auto-correlation function of
the off-diagonal terms of pressure tensor as given in Eq.~\ref{eq:acf}. Since
correlation of pressure tensor only depends on the number of data points, rather than
system size, these calculations are performed only for the system
containing 6720 water molecules. For the computation, a total of
$2\times10^6$ data 
points are considered with correlation time corresponding to $5\times10^5$ data
points. The corresponding logic for computing viscosity is described in
Algorithm~\ref{al:acf}.

\begin{equation}
 \label{eq:acf}
\eta(dt) = \frac {V}{k_BT} \int_0^\infty dt \langle {\bf P}_{\alpha \beta}(t).{\bf P}_{\alpha \beta}(t+dt) \rangle
\end{equation}
\begin{lstlisting}[frame=lines, language=Fortran, mathescape,
caption={Serial logic for auto-correlation function}, label={al:acf}]
Do dt = 0 to Nframe-1         ! Loop 1 over time interval
   Do t = 1 to Nframe-dt      ! Loop 2 over number of frames
      Compute product of pressures at different times
      Update correlation array
   End Do                     ! End of loop 2
End Do                        ! End of loop 1
Do dt = 0 to Nframe-1         ! Loop 3 over time interval
   Perform integration using quadrature method for Eq.$~\ref{eq:acf}$
End Do                        ! End of loop 3  
\end{lstlisting}

Since all computations for the product of pressure are independent, we parallelize over
correlation times (dt) as shown in Algorithm~\ref{al:acfpar}.
\begin{lstlisting}[frame=lines, language=Fortran,
mathescape,caption={Parallel logic for auto-correlation function}, label={al:acfpar}]
Allocate memory for time data and correlation array on CPU and GPU
Transfer time data to GPU
call CUDA_acf
Transfer correlation array to CPU
Do dt = 0 to Nframe-1         ! Loop 3 over time interval
   Perform integration using quadrature method for Eq.$~\ref{eq:acf}$
End Do                        ! End of loop 3  
free memory

function CUDA_acf
    dt is the thread id
    Do t = 1 to Nframe-dt      ! Loop 1 over number of frames
      Compute product of pressures at different times
      Update correlation array
    End Do 
End function
\end{lstlisting}
Using this strategy, the computing times obtained for CPU, GTX, and K40 are
1534 s, 76.2 s, and  84.1 s which indicate the speed up of $\approx
18-20\times$ of GPUs over CPU.

\subsubsection{Mean Square Displacement}
Mean square displacement (MSD) averaged over all the atoms can be used to gain the information about the
self diffusion behaviour of a particle in a given media. To obtain self-diffusion constant, Einstein relation can be 
utilized which is given as\cite{fs02} 
\begin{equation}
D = \lim _{t\to\infty} \frac {1}{2dt} \langle |{\bf r}(t) - {\bf r}(0)|^2 \rangle,
\label{eq:msd}
\end{equation}
where $d$ is the dimensionality of the system, while 
$\langle \cdots \rangle$ denotes the MSD with averaging done over all
the atoms (or all the atoms in a given subclass), to improve statistical
accuracy. The pseudocode for  serial calculation is given in Algorithm~\ref{al:msd}.
As shown in Algorithm~\ref{al:msd}, MSD involves looping 
over all atoms, all time origins and correlation times.

\begin{lstlisting}[frame=lines, language=Fortran, mathescape, caption={Serial logic for mean square displacement}, label={al:msd}]
!Starting from unwrapped trajectory
Do i = 1 to Natom                ! Loop 1 over the particles
   Do dt = 0 to Nframe-1         ! Loop 2 over time interval
      Do t = 1 to Nframe-dt      ! Loop 3 over number of frames
         Compute MSD of particle i between t and t+dt
         Accumulate MSD array
      End Do                     ! End of Loop 1  
   End Do                        ! End of Loop 2
End Do                           ! End of Loop 3

\end{lstlisting}
Just like the ACF described above (Algorithm~\ref{al:acf}), MSD can also be
parallelized over the correlation time (dt). Since, unlike ACF, MSD also depends on 
system size, it can be, additionally,
parallelized over all atoms. This also affords division of data over
multiple GPUs, if required. We describe this in Algorithm~\ref{al:msdpar}.

\begin{lstlisting}[frame=lines, language=Fortran, mathescape, caption={Parallel logic for mean square displacement}, label={al:msdpar}]
!Starting from unwrapped trajectory
Allocate memory for arrays of positions and msd on CPU and GPU
Transfer unwrapped position arrays to GPU
Call CUDA_msd
Transfer msd array from GPU
Free memory

function CUDA_msd
dt and atom index is obtained from thread-id
    Do t = 1 to Nframe-dt      ! Loop 1 over number of frames
       Compute MSD of particle i between t and t+dt
       Accumulate local MSD data
    End Do                     ! End of Loop 1  
    Update global MSD array
End function
\end{lstlisting}
\begin{table}[htbp]
 \centering
\begin{tabular}{@{\extracolsep{4pt}}rrrrrr@{}}
\hline 
\hline 
\multicolumn{1}{c}{System Size} & \multicolumn{1}{c}{CPU} & \multicolumn{2}{c}{GTX} & \multicolumn{2}{c}{K40} \\ 
\cline{3-4}
\cline{5-6}
 & Time (s) & Time (s) & Speed up & Time (s) &Speed up\\
\hline 
 6720  &  664 & 8   & 83 & 10 & 66 \\ 
 13440 & 1310 & 15  & 87 & 21 & 62 \\ 
 26880 & 2600 & 31  & 83 & 43 & 40 \\ 
\hline
\end{tabular}
\caption{Performance comparisons for computation of mean square displacement
 using conventional serial CPU and parallel CUDA codes, on GTX and K40
 machines. Computation is performed for 3500 frames and relative speed up is
 also given. Note that time of file reading is not included in CPU time.}
\label{tab:msd}
\end{table}
MSD computation is among the most memory intensive calculation in this
study and hence is considerably aided by splitting data sets over atoms. On
a single GPU, speed up of $\approx80\times$ are easily achieved for systems
with $\approx27000$ particles, as shown in Table~\ref{tab:msd}, i.e. from
43 minutes to 31 {\it seconds} on a GTX780 for 27000 particles. {\it n.b.}
3500 frames were chosen so as to completely utilize memory of a GTX780
card. The speedup is nearly consistant for the GTX card with 80x speedup for the smallest system as well. In comparison, time taken for the same calculation for 6720 particles on a V100 card is 5.9 s. The time saved by such an implementation is evident in computations where
there are a larger number of state points, e.g. diffusivity calculations for
more than 300 state points for $n$-alkanes in Ref \citenum{sack17}. Mean square displacement calculation is common enough that software suites such as DL\_POLY, GROMACS etc have their own optimized CPU codes to compute it. GROMACS MSD CPU code (gmx msd) takes $\approx$171 s for the calculation of MSD for 3500 frames for 6720 particles (including time taken for reading the data). Hence, the resulting speedup for the CUDA code running on K40, for example, would be $\approx$17. Considering that the unoptimized GPU code provides a much faster performance than an optimized serial CPU only code, besides being a completely in-house code, it would be useful to implement CUDA driven code for such embarrassingly parallel problems. 

\subsection{Parallelization over CPUs \label{ssec:cpuparallel}}
In the case of trajectory analyses algorithms, it is possible to devise shell script
wrappers for serial codes which would make the calculation parallel over the
available hardware. For example, a serial code can be converted into a parallel
code over $n$ cores by breaking the trajectory into multiple sets and using a
wrapper script or program to control the execution. A separate program would
be needed to collate the results from the multiple serial runs. Assuming near
100\% efficiency and ignoring file reading/writing overheads, the best speed up
with this method would be $n\times$. For example, in this study, the best speed
up we can expect is 24$\times$ for the K40 machine and 6$\times$ for the GTX
machine. Considering that each machine has 2 and 1 GPUs respectively, a similar
GPU speed up, e.g. for computation of MSD, would be $\approx80\times$ (on both
machines). It must be noted that if a CPU parallel scheme is used, the entire
machine would be unavailable for other programs. On the other hand, if a GPU
code is used, the other CPU cores would be available for 
processes. As shown in the results, there are very few cases where the GPU speed
up over the serial code is less than $10\times$, and hence would be more
efficient than using a parallel CPU code. 

Studies involving large number of state points, for example Ref
\citenum{sack17}, the combined time saved is of considerable importance,
considering the cost of high performance computing. In Ref \citenum{sack17},
$\approx$300 state points have been simulated, and $g^{(3)}$ was computed with
5000 frames for each state point in addition to RDFs and MSDs. With a conventional CPU code, the estimated $g^{(3)}$
computation time would be $\approx$43-44 days parallelized over 24 cores (see
Table \ref{tab:g3}). In comparison, the task would be completed in $\approx$28
days on 2 K40 GPUs, or $\approx$16-17 days on 2 GTX 780 GPUs including time taken for all file
I/O operations. 

\section{Conclusion}
MD simulations are routinely used as an investigating tool for range of systems.
Large system sizes in MD simulations result in large trajectories which demand
swift analyses of multiple varieties. Like the simulations themselves, the
analyses too have some parallelizable regions which can be easily implemented on
NVIDIA GPUs, and hence it is imperative to port these codes from CPU to GPU.
Multiple algorithms, with low to moderate complexities, are described with ease
of implementation in mind. Although the corresponding OpenMP and MPI based
algorithms can be coded, the speed up demonstrated above can be achieved only
after significant optimization and code modification. Even though the usage cost
of GPUs is $\approx10\times$ that of a single CPU core in a typical commercial
HPC setup, the effective cost of analysis on GPU would be considerably lower
than that on CPUs. Since these codes do not use any advanced GPU architecture
based functionality, they would be compatible with future generations of NVIDIA
devices, as demonstrated on V100 GPUs. 

Static properties such as $q_{tet}$, $P_2$, and $g^{(3)}$ can be easily
parallelized over atoms as well as frames while the trajectory can be
partitioned to conform to memory limitations and make use of multiple devices.
Using the approaches described in this work, the process of discovery of both
pairs and triplets can be accelerated significantly. For dynamic
quantities, such as MSD, all trajectory data for a single species are required
to be present on the device and, therefore, data is needed to be divided over
specific atoms or species.  Hence, parallelization can be done over trajectory
frames or atoms as per requirement.  Typical speed up obtained are 10-80$\times$
over conventional CPU codes, allowing researchers to quickly analyze
data on both server class NVIDIA GPUs as well as Desktop and workstation level
consumer GPUs. Since similar approaches such as auto-correlation functions and
neighbour search are also used in other scientific fields, these algorithms
can be easily extended. For analyses such as MSD and $g^{(3)}$ which require considerably large memory, and are data heavy computations, the code must be ideally parallelized further using distributed memory approaches such as MPI, as well as make use of the CPUs of the machines being used. In future, we plan to update the repository with other examples of analysis codes using CUDA, as well as directive based programming standards such as OpenACC.

\section*{Acknowledgement} 
This paper is dedicated to the memory of Late Prof. Charusita Chakravarty. The
work described above started in January 2014 with the aim of reducing time taken
for computation of three particle correlation function in the group.  G.S.
thanks the CSIR, India for granting the Senior Research Fellowship (F. No. -
09/086(1141)/2012-EMR-I).  Authors thank the IIT Delhi HPC facility for
computational resources. Authors appreciate discussions with NVIDIA development
team. Authors also thank C-DAC and NVIDIA for providing access to P100 and V100 cards on the `Sangam' and `Sreshta' testbeds during the NVIDIA Hackathon (17th to 21st September 2018) for the calculations.

\bibliographystyle{elsarticle-num}
\bibliography{ref-gpu-methods}
\end{document}